\documentclass{aa}
\usepackage{graphicx}
\begin{document}
   	\title{Photometric variability of young brown dwarfs in the $\sigma$
	Orionis open cluster}


   	\author{J. A. Caballero\inst{1}
          	\and
   	  	V. J. S. B\'ejar\inst{1}
          	\and
   	  	R. Rebolo\inst{1,2}
          	\and
          	M. R. Zapatero Osorio\inst{3}}

   	\offprints{J. A. Caballero}

   	\institute{Instituto de Astrof\'{\i}sica de Canarias, E-38205 La
	Laguna, Tenerife, Spain\\
        \email{zvezda@ll.iac.es}
        \and
        Consejo Superior de Investigaciones Cient\'{\i}ficas, Spain
        \and
        LAEFF-INTA, P.O. Box 50727, E-28080, Madrid, Spain}

   	\date{Received January 09, 2004; accepted May 28, 2004}

   	\abstract{We have carried out multi-epoch, time-series
	differential $I$-band photometry of a large sample of objects
	in the south-east region of the young ($\sim$3\,Myr), nearby
	($\sim$350\,pc) $\sigma$ Orionis open cluster.  A field of
	$\sim$1000\,arcmin$^2$ was monitored during four nights over a
	period of two years.  Using this dataset, we have studied the
	photometric variability of twenty-eight brown dwarf
	cluster candidates with masses ranging from the
	stellar--substellar boundary down to the planetary-mass
	domain.  We have found that about 50\%~of the sample
	show photometric variability on timescales from less than one
	hour to several days and years.  The amplitudes of the
	$I$-band light curves range from less than 0.01 up to $\sim$0.4
	magnitudes.  A correlation between the near-infrared excess in
	the $K_{\rm s}$ band, strong H$\alpha$ emission and
	large-amplitude photometric variation is observed.  We briefly
	discuss how these results may fit the different scenarios
	proposed to explain the variability of cool and ultracool
	dwarfs (i.e.\ magnetic spots, patchy obscuration by dust
	clouds, surrounding accretion discs and binarity).
	Additionally, we have determined tentative rotational periods
	in the range 3 to 40\,h for three objects with masses around
	60\,$M_{\rm Jup}$, and the rotational velocity of 14 $\pm$
	4\,km\,s$^{-1}$ for one of them.  }
	
	\titlerunning{Variability of young brown dwarfs in $\sigma$ Orionis}

   	\maketitle
%

\section{Introduction}

The photometric variability of brown dwarfs and very low mass stars
close to the substellar limit has received significant attention
during the past few years.  The studies, mostly performed in the red
part of the optical spectrum, using broad-band (Mart\'{\i}n \&
Zapatero Osorio 1997; Terndrup et al$.$ 1999; Bailer-Jones \& Mundt
1999, 2001; Mart\'{\i}n, Zapatero Osorio \& Lehto 2001; Gelino et al$.$
2002; Joergens et al$.$ 2003) and narrow-band filters (Tinney \& Tolley
1999; Clarke, Tinney \& Covey 2002; Clarke, Oppenheimer \& Tinney
2002), indicate that approximately one third of the late M- and L-type
dwarfs show variability with amplitudes from $\sim$10 to $\sim$80
milimagnitudes (mmag).  Several of these objects have been reported as
periodic variables, with periods in the range from half an hour to ten
days.  Photometry in the near-infrared region, where cool dwarfs emit
most of their output energy, has also been performed in young clusters
(Carpenter et al$.$ 2002; Zapatero Osorio et al$.$ 2003---hereafter ZO03)
and in the field (Bailer-Jones \& Lamm 2003; Enoch, Brown \& Burgasser
2003).  The detected amplitudes of variation are in the range
0.05--0.2\,mag, and tentative periods of $\sim$1.5 to 3\,h have been
reported.  On the other hand, time-resolved spectroscopy has
emphasized the study of dust/temperature- or activity-sensitive
spectral features such as water vapour, TiO, FeH and CrH bands or
H$\alpha$ emission, respectively (Nakajima et al$.$ 2000; Hall 2002;
Bailer-Jones 2002; Burgasser et al$.$\ 2002; Liebert et al$.$ 2003;
Clarke, Tinney \& Hodgkin 2003).  Among the scenarios proposed to
explain the observed levels of photometric and spectroscopic
variability, we may list cool and/or hot corotating
magnetically-induced spots in the atmospheres, heterogeneous or patchy
coverage of photospheric clouds of grains of solid condensates,
(sub)stellar discs surrounding the objects, and the presence of very
low-mass companions in close orbits, which may produce eclipses or
mass-transfer episodes.

Photometric variability studies of objects with confirmed
membership in star clusters with known astrophysical properties have advantages
to those in the field.   
Age, metallicity and distance are the same for the
sample of cluster members. Because they lie within a
well-defined photometric sequence, different luminosities correspond
to different masses. Hence, the monitoring of a sample of cluster
objects usually leads to the exploration of the properties of members
in a wide range of masses. Additionally, wide field cameras can
simultaneously observe a relatively large number of sources under the
same sky and instrumental conditions. 
At very young ages (e.g., a few
Myr), fast-rotating completely convective brown dwarfs may generate a
magnetic field strong enough to produce solar-like surface spots or
chromospheric activity.  
Also, accretion discs may not yet have been completely dissipated.
Investigation of the rotation of very young brown dwarfs is especially
important for understanding crucial phenomena related to substellar
formation, such as angular momentum evolution, the relation between
disc and central object, and disc evaporation.

Near-infrared and optical wide-field photometric monitoring of stars
in the Orion region has already been performed, especially in the area
surrounding the Orion Nebula Cluster (Herbst et al$.$ 2000, 2002;
Carpenter, Hillenbrand \& Skrutskie 2001; Rebull 2001 and references
therein).  In particular, in the $\sigma$ Orionis region, evidence of
variability has been found in a large variety of objects, from the
massive magnetic helium-strong B2Vp star in the central OB multiple
stellar system, through solar-like stars, to objects below the
stellar--substellar threshold (e.g.\ Fedorovich 1960; Landstreet \&
Borra 1978; Bailer-Jones \& Mundt 2001).

In this paper we present differential photometry of 32 low-mass member
candidates of the young $\sigma$ Orionis open cluster, which belongs
to the Ori OB1b association.  Walraven photometry and {\sc hipparcos}
measurements of the distance to the brightest stars of the association
point to a distance of $\sim$350\,pc (Brown, de Geus \& de Zeeuw 1994;
Perryman et al$.$ 1997; de Zeeuw et al$.$ 1999).  Optical extinction
towards the $\sigma$ Orionis stellar system and the surrounding region
seems to be small ($A_V <$1\,mag---Lee 1968; B\'ejar et al$.$\ 2001;
Oliveira et al$.$\ 2002).  The age of the cluster is estimated between 1
and 8\,Myr, with 3\,Myr being the most likely value (Zapatero Osorio
et al$.$\ 2002a; Oliveira et al$.$\ 2002).

\section{Observations and analysis}

The observations were performed at Observatorio del Roque de Los
Muchachos, on La Palma (Canary Islands).  We used the four-chip Wide
Field Camera (WFC) mosaic, mounted at the prime focus of the 2.5\,m
Isaac Newton Telescope.  The detector consists of four thinned EEV4280
2k $\times$ 4k CCDs with a pixel size of 13.5\,$\mu$m, corresponding
to a pixel scale of 0.333\arcsec.

Monitoring of a $\sim$0.29\,degree$^2$ area south-east of the centre
of the $\sigma$ Orionis cluster was carried out during two epochs:
three consecutive nights, from 2000 December 30 through 2001 January 1
(hereafter, WFC00), and on 2003 January 8 (WFC03).  Several Landolt
standard stars (Landolt 1992) were observed during 2003 January 8,
which allowed us to calibrate the WFC03 $I$ magnitude.  Thirty-three
different exposures, each with completeness magnitude $I\sim$21.5,
resulted from these observations.  We used the RGO $I$ filter, which
resembles the Johnson--Cousins--Kron $I$.  Although the nights were
clear, seeing conditions varied slightly during the runs.  In the
first epoch, a longer exposure time was used, and the counts of
several bright objects lay in the non-linear regime when the seeing
improved.  In Table \ref{log} is a log of the observations with other
useful information.  We give the date, number of exposures taken each
observing night, effective mean airmass, average full width half
maximum (FWHM) in arcsec and exposure time in seconds.  Above the
horizontal line is shown the WFC00 dataset, and below is the WFC03
dataset.  The WFC00 data have already been used, together with
near-infrared observations, to study the photometric variability of
the young, low-mass brown dwarf S Ori 45 (ZO03).

\begin{table}
    \centering
    	\caption[]{Log of the observations}  
        \label{log}
        \begin{tabular}{lcccc}
            	\hline
            	\noalign{\smallskip}
Date 		& No.  		& Airmass	& FWHM 		& Exp.\ time \\	
		& exposures 	& 	& [\arcsec] 	& [s] \\	 
            	\noalign{\smallskip}
            	\hline
            	\noalign{\smallskip}
30 Dec 2000   	& 5 		& 1.23--1.70  	& 1.4--1.6	& 1500 \\
31 Dec 2000   	& 10  		& 1.17--1.50  	& 0.9--1.4	& 1500 \\
01 Jan 2001   	& 6 		& 1.17--1.58  	& 1.2--1.8	& 1500 \\
            	\noalign{\smallskip}
            	\hline
            	\noalign{\smallskip}
08 Jan 2003   	& 12  		& 1.17--1.81	& 1.0--2.0	& 1200 \\
           	\noalign{\smallskip}
            	\hline
         \end{tabular}
\end{table}

The reduction of the images was performed within the {\sc
iraf}\footnote{{\sc iraf} is distributed by National Optical Astronomy
Observatories, which are operated by the Association of Universities
for Research in Astronomy, Inc., under cooperative agreement with the
National Science Foundation.}  environment, using standard packages
({\sc imred--ccdred}).  Raw images were bias-subtracted and
flat-fielded before alignment.  A dither pattern, with displacements
of several times the seeing, was used in order to subtract the
contribution of the fringing appropriately.  To accomplish this, a
flat-field image for each night was constructed using bias-subtracted
scientific images.
 
Alignments of the WFC00 and WFC03 images were done separately, because
of the presence of a 0.6-arcmin shift of pointing centres between
epochs.  Residual fringing was below the level of 0.8\,\% over the
CCDs.  A study of the average FWHM and sky background of each image
was made before calculating the aperture photometry.  This analysis
was done for each detector individually.  We used the {\sc iraf} {\sc
daophot} package to perform the photometric analysis.  Aperture
photometry was performed using nine different apertures per image:
0.50, 0.60, 0.70, 0.80, 0.90, 1.00, 1.25, 1.50, and 2.00 times the
average FWHM.  The inner radius of the sky annulus was 5.00 times the
average FWHM, while the width of the annulus was fixed at 8 pixels.
Instrumental magnitudes for 29514 objects were obtained.

\subsection{Differential photometry}

The data analysis (selection of sources, calculation of the reference
magnitudes, statistical analysis, obtaining of light curves,
preliminary time-series analysis) was done mainly with the {\sc
matlab}\footnote{{\sc matlab} is a high-performance language for
technical computing developed by The MathWorks, Inc.}  interactive
system.

We required the mean of the {\sc iraf} magnitude errors from sources
of the whole sample to be lower than 0.25\,mag for each of the nine
apertures.  In order to avoid source confusion, especially when faint
objects are close to bright ones or in crowded fields, objects with
standard deviations of the centroid coordinates larger than 1.5 pixels
($\sim$0.5\arcsec) were rejected.  Taking into account these
requirements, sources with counts close to saturation, very faint
objects that appear close to very bright stars, cosmic rays and chip
defects were mostly discarded.

Instrumental magnitudes were transformed into differential
magnitudes, $I_{\rm diff}$, by subtracting a reference value (i.e. a
zero-point magnitude level). 
This reference magnitude for each CCD
and exposure was calculated using a large number of bright,
non-variable stars located over the entire monitored area.  They were
chosen using an iterative procedure: only low-photometric-noise
sources with standard deviation $\sigma(I_{\rm diff})<$ 0.020\,mag (in
the WFC00 dataset) and $<$ 0.015\,mag (WFC03), and with magnitudes in
the range 17.5\,mag $<I<$ 21.0\,mag, were selected from an
initial sample of reference star candidates.  Each CCD of the WFC00
and WFC03 datasets was treated separately.  The final sample of
reference objects ($\sim$100 per chip) was the intersection between
the WFC00 and WFC03 reference samples. We note that a photometrically
stable brown-dwarf target, when compared to other objects of similar
magnitude, may also be considered a reference ``star'' if it matches
the criteria given above.

Reference magnitudes were calculated following Bailer-Jones \& Mundt (2001).  
The method is based in a mean of fluxes of the reference
stars, where the brightest sources, with the lowest Poissonian
photon-count errors, dominate the final reference magnitude.  However,
a simple mean of the magnitudes of the reference stars was also
worked out, to check for any significant difference.  Because of the
large size of the sample of reference stars, both zero-point
magnitudes were similar.  As the standard deviation of the difference
between the mean of fluxes and the mean of magnitudes was lower than
1\,mmag, we used only the mean-of-fluxes reference level in the
subsequent calculations.

We studied the dependence of the reference magnitude for each CCD on
time, airmass and seeing.  No trends were detected, so we did not
applied either airmass- or seeing-dependent high-order corrections.
Computations were made for each Wide Field Camera chip separately, as
each CCD has different characteristics (i.e. gain, electronic noise
and offsets between instrumental magnitudes and apparent magnitudes).
Also, for each chip the average FWHM is different at a given time
because of the different distances to the centre of the focal plane.

We performed the previous analysis in the WFC00 dataset for each of
the nine apertures (0.5 to 2.0 $\times$ FWHM) and chose the 0.70
$\times$ FWHM aperture, which both minimized the photon contribution
of nearby contaminant sources and maximized the signal-to-noise ratio
and, as a result, provided the largest number of reference stars.

An additional offset was added to the $I_{\rm diff}$ values of each
object in order to transform them into apparent magnitudes, calibrated
using the Landolt standard stars observed during the WFC03 run.  As a
result, for each object the mean WFC03 $I_{\rm diff}$, denoted by
$\overline{I_{03}}$, is equal to the real $I$-band magnitude.
 
Finally, multi-epoch calibrated $I$-band light curves were obtained
for more than 8000 objects that fulfilled all the criteria given
above.

\section{Substellar candidates}

   \begin{figure}
   \centering
   \includegraphics[width=0.5\textwidth]{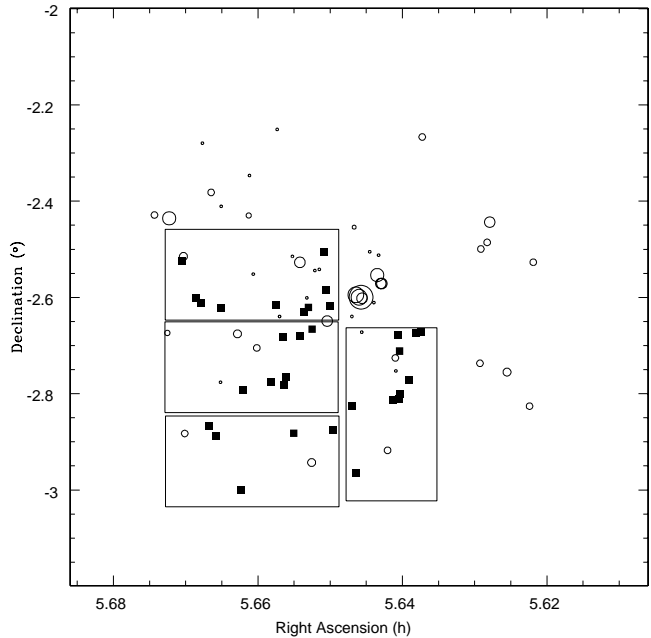}
      \caption{Spatial location of the 32 brown-dwarf member candidates in the
      $\sigma$ Orionis cluster studied in this article, shown with filled squares.
      Stars brighter than $I$ = 15.0\,mag that presumably belong to the cluster are
      represented with open circles. 
      The circle radii are inversely proportional to the magnitude.
      Rectangles denote the four CCD chips of the Wide Field Camera.}
         \label{fso_dec_ra}
   \end{figure}

   \begin{figure}
   \centering
   \includegraphics[width=0.5\textwidth]{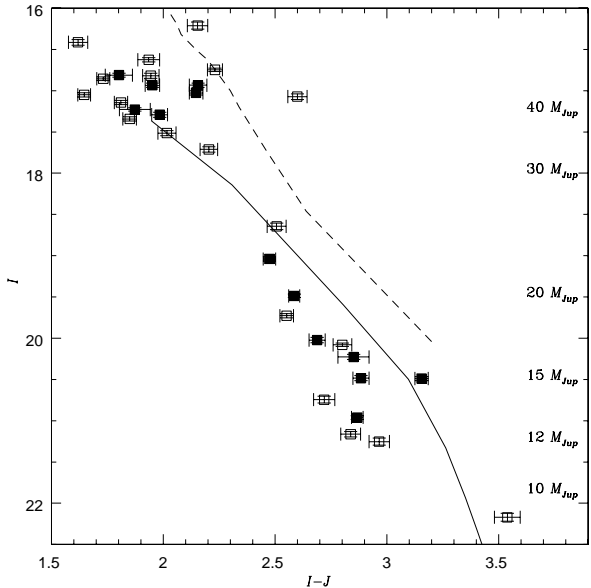}
      \caption{$I$ vs.\ $I-J$ colour--magnitude diagram of the 32 brown-dwarf
      	member candidates under study.
      	Spectroscopically confirmed substellar cluster members are shown with
      	filled squares, while open squares indicate objects with no available
	spectroscopic data.
      	The solid and dashed lines are 3 Myr colour--magnitude isochrones for
	substellar objects, based on models by the Lyon group (Baraffe et al$.$
	2002 and Chabrier et al$.$ 2000, respectively).  
	Approximate masses are provided to the right.} 
         \label{IIJ}
   \end{figure}

\begin{table*}
     \centering
    	\caption[]{Photometric and spectroscopic data for the 32 $\sigma$
	Orionis cluster members considered in our study}   
        \label{spectrum}
        \begin{tabular}{lcccccccc}
            	\hline
            	\noalign{\smallskip}
Name & $I$ & $I-J$ & Sp.\ type & pEW(Li {\sc i}) (\AA) & 
\multicolumn{4}{c}{pEW(H$\alpha$) (\AA)} \\
            	\noalign{\smallskip}
		\cline{6-9}
            	\noalign{\smallskip}
 & & &  & & BZOR & ZO & ByN & M \\	  
            	\noalign{\smallskip}
		\hline
            	\noalign{\smallskip}
\object{S Ori J054000.2$-$025159} 		& 16.21$\pm$0.04 	& 2.15$\pm$0.05 	& --	& --	& --	& --	& --	& --	\\ 
\object{S Ori J053902.1$-$023501} 		& 16.41$\pm$0.03 	& 1.62$\pm$0.05 	& --	& --	& --	& --	& --	& --	\\ 
\object{S Ori 16}$^{\mathrm{a}}$		& 16.62$\pm$0.03 	& 1.94$\pm$0.04 	& --	& --	& --	& --	& --	& --	\\ 
\object{S Ori J053847.2$-$025756} 		& 16.75$\pm$0.02 	& 2.23$\pm$0.03 	& --	& --	& --	& --	& --	& --	\\ 
\object{S Ori J053829.0$-$024847}$^{\mathrm{a}}$ & 16.81$\pm$0.02 	& 1.80$\pm$0.06 	& M6.0$\pm$0.5 	& --	& --	& --	& $<$ 5 & \\ 
\object{S Ori J053954.3$-$023719}$^{\mathrm{a}}$ & 16.82$\pm$0.03 	& 1.94$\pm$0.04 	& --	& --	& --	& --	& --	& --	\\ 
\object{S Ori J053825.4$-$024241}$^{\mathrm{a}}$ & 16.86$\pm$0.02 & 1.73$\pm$0.03 	& --	& --	& --	& --	& --	& --	\\ 
\object{S Ori 25}$^{\mathrm{a}}$ 		& 16.93$\pm$0.03 	& 2.16$\pm$0.03 	& M7.5$\pm$0.5 	& 0.6 & 45.0$\pm$1.0 	& --	& 42$\pm$8 & 44$\pm$1 \\ 
\object{S Ori J053826.1$-$024041}$^{\mathrm{a}}$ & 16.93$\pm$0.02 	& 1.95$\pm$0.04 	& M8.0$\pm$0.5 	& --	& --	& --	& 4$\pm$2 & --	\\ 
\object{S Ori 27}$^{\mathrm{a}}$ 		& 17.03$\pm$0.03 	& 2.15$\pm$0.03 	& M7.0$\pm$0.5 	& 0.74$\pm$0.09 & 6.1$\pm$1.0 & 5.7$\pm$0.5 & 5$\pm$2 & --	\\ 
\object{S Ori J053922.2$-$024552} 		& 17.05$\pm$0.03 	& 1.65$\pm$0.05 	& --	& --	& --	& --	& --	& --	\\ 
\object{S Ori J054014.0$-$023127} 		& 17.07$\pm$0.03 	& 2.60$\pm$0.04 	& --	& --	& --	& --	& --	& --	\\ 
\object{S Ori 28}$^{\mathrm{a}}$		& 17.14$\pm$0.03 	& 1.81$\pm$0.03 	& --	& --	& --	& --	& --	& --	\\ 
\object{S Ori 31}$^{\mathrm{a}}$ 		& 17.23$\pm$0.02	& 1.87$\pm$0.07	  	& M7.0$\pm$0.5 	& --	& --	& --	& 2.5$\pm$0.9 & --	\\ 
\object{S Ori 30}$^{\mathrm{a}}$ 		& 17.29$\pm$0.03	& 1.98$\pm$0.03		& M6.0$\pm$0.5 	& --	& --	& --	& 16$\pm$6 & --	\\ 
\object{S Ori 32}$^{\mathrm{a}}$		& 17.34$\pm$0.03 	& 1.85$\pm$0.03 	& --	& --	& --	& --	& --	& --	\\ 
\object{S Ori J054004.5$-$023642}$^{\mathrm{a}}$ & 17.51$\pm$0.03 	& 2.02$\pm$0.04 	& --	& --	& --	& --	& --	& --	\\ 
\object{S Ori 36}$^{\mathrm{a}}$		& 17.71$\pm$0.03 	& 2.20$\pm$0.04 	& --	& --	& --	& --	& --	& --	\\ 
\object{S Ori J053918.1$-$025257} 		& 18.64$\pm$0.04 	& 2.51$\pm$0.04 	& --	& --	& --	& --	& --	& --	\\ 
\object{S Ori 42}$^{\mathrm{a}}$ 		& 19.04$\pm$0.03	& 2.48$\pm$0.03		& M7.5$\pm$0.5 	& --	& --	& --	& 89$\pm$12 & --	\\ 
\object{S Ori 45}$^{\mathrm{a}}$ 		& 19.49$\pm$0.02	& 2.57$\pm$0.02		& M8.5$\pm$0.5 	& 2.4$\pm$1.0 & 60.0$\pm$1.0 & 33$\pm$9 & 26$\pm$15 & --	\\ 
\object{S Ori J053929.4$-$024636} 		& 19.73$\pm$0.03 	& 2.55$\pm$0.03 	& --	& --	& --	& --	& --	& --	\\ 
\object{S Ori 71}$^{\mathrm{a}}$ 		& 20.02$\pm$0.03 	& 2.69$\pm$0.04 	& L0.0$\pm$0.5 	& --	& --	& --	& 700$\pm$80 & --	\\ 
\object{S Ori J053849.5$-$024934} 		& 20.08$\pm$0.02 	& 2.80$\pm$0.04 	& --	& --	& --	& --	& --	& --	\\ 
\object{S Ori 51}$^{\mathrm{a}}$ 		& 20.23$\pm$0.03 	& 2.85$\pm$0.07 	& M9.0$\pm$0.5 	& --	& --	& --	& 25: & --	\\ 
\object{S Ori 50}$^{\mathrm{a}}$ 		& 20.48$\pm$0.03 	& 2.88$\pm$0.04 	& M9.0$\pm$0.5 	& --	& --	& --	& $<$ 10 & --	\\ 
\object{S Ori 47}$^{\mathrm{a}}$ 		& 20.49$\pm$0.02 	& 3.16$\pm$0.03 	& L1.5$\pm$1.0 	& 4.3$\pm$0.5 	& --	& $\le$ 6 & 25: & --	\\ 
\object{S Ori J053944.5$-$025959} 		& 20.74$\pm$0.04 	& 2.72$\pm$0.05 	& --	& --	& --	& --	& --	& --	\\ 
\object{S Ori 53}$^{\mathrm{a}}$ 		& 20.96$\pm$0.02 	& 2.87$\pm$0.03 	& M9.0$\pm$0.5 	& --	& --	& --	& $<$ 10 & --	\\ 
\object{S Ori J054007.0$-$023604}		& 21.16$\pm$0.03 	& 2.84$\pm$0.04 	& --	& --	& --	& --	& --	& --	\\ 
\object{S Ori J053956.8$-$025315} 		& 21.25$\pm$0.04 	& 2.97$\pm$0.05 	& --	& --	& --	& --	& --	& --	\\ 
\object{S Ori J053858.6$-$025228}		& 22.17$\pm$0.05 	& 3.54$\pm$0.06 	& --	& --	& --	& --	& --	& --	\\ 
        	\noalign{\smallskip}
            	\hline
         \end{tabular}
	\begin{list}{}{}
	\item [$^{\mathrm{a}}$] References: B\'ejar, Zapatero Osorio \& Rebolo
	1999 and B\'ejar et al$.$ 2004 (BZOR); Zapatero Osorio et al$.$
	1999, 2002a (ZO); Barrado y Navascu\'es et al$.$ 2001, 2002, 2003 (ByN);
	Muzerolle et al$.$ 2003 (M). 
	\end{list}
\end{table*}

Of the 32 $\sigma$\,Orionis brown dwarf candidates studied here, 19
were photometrically discovered at optical wavelengths by B\'ejar et
al. (1999, 2004).  Their near-infrared colours are consistent with
membership of the cluster.  In addition, many of them have been
confirmed spectroscopically.  The remaining 13 objects were selected
from a combined $IJ$ survey conducted with the Wide Field Camera ($I$)
at the Isaac Newton Telescope and the {\sc isaac} instrument ($J$) at
the Very Large Telescope Antu/UT 1.  Details of this survey will be
given in a forthcoming paper.  The 32 selected targets for the
present photometric analysis have $I$ magnitudes between 16.0 and
22.5\,mag.  
Neither brighter nor fainter objects yielded reliable
differential photometry either because of non-linearity or excessively
faint signals in each exposure, respectively.  In
Fig. \ref{fso_dec_ra}, we show the spatial location of the 
targets, stars brighter than $I$ = 15.0\,mag that presumably belong
to the $\sigma$ Orionis cluster (based on spectral type and/or proper
motion determinations) and the position on the sky of the WFC mosaic.
We provide in Table \ref{spectrum} the complete sample, together
with relevant spectroscopic information from the literature (e.g.,
spectral types and pseudo-equivalent widths of lithium absorption at
6708\,\AA~ and H$\alpha$ emission).

The spectral types of the targets range from $\sim$M5.5 to $\sim$L2.
According to theoretical models, at the age of the cluster
($\sim$3\,Myr), these roughly correspond to masses\footnote{Conversion
between solar and Jupiter masses: 1\,$M_{\odot}$ = 1047.56 $\pm$
0.08\,$M_{\rm Jup}$.}  between $\sim$80 and $\sim$8\,$M_{\rm Jup}$.
The upper limit is slightly larger than the theoretical minimum mass
for hydrogen fusion, the borderline between stars and brown dwarfs.
The lower limit is below the deuterium-burning mass threshold,
suggested as the frontier between brown dwarfs and planetary-mass
objects (e.g.\ Saumon et al$.$\ 1996).  These mass estimates may be
affected by uncertainties of the evolutionary models at very young
ages (see the discussion by Baraffe et al$.$ 2002, 2003).  The
spectrophotometric sequence of the $\sigma$ Orionis cluster in the
substellar domain is known from previous studies (references in
footnote of Table \ref{spectrum} and B\'ejar et al$.$ 2001).  A
colour--magnitude diagram $I$ vs. $I-J$, such as the one shown in Fig.
\ref{IIJ}, provides a reliable way of determining the membership of
candidates without spectroscopic data.

\subsection{H$\alpha$ emission}

According to the standard scenario of classical T Tauri stars,
broad H$\alpha$ emission lines are caused by the accretion flow of
free-falling gas.  The infalling gas, in this case, is accreted during
the early stages of star formation on to the disc or at the base of
high-latitude magnetic accretion columns on the stellar surface
(K\"onigl 1991, and references therein).  Several authors have
suggested the extrapolation of this picture to objects below the
stellar/substellar mass limit.

The great majority of the spectroscopically confirmed brown dwarfs
show H$\alpha$ emission (see Table~\ref{spectrum}), which may be
related to chromospheric activity or to the presence of a
circum(sub)stellar disc.  The asymmetric, extraordinarily strong
H$\alpha$ emission of the cool substellar object S Ori 71 should be
noted.  Barrado y Navascu\'es et al$.$ (2002) estimated its mass to be
between 22 and 13\,$M_{\rm Jup}$.  They measured a pseudo-equivalent
width of 700 $\pm$ 80\,\AA, which is among the strongest H$\alpha$
emissions so far seen in a young late-type dwarf.  S Ori 25, S Ori 42
and S Ori 45 also show strong H$\alpha$ emission, which is
$\sim$10--20 times greater than the average emission of other $\sigma$
Orionis sources of similar spectral type.  According to Barrado y
Navascu\'es \& Mart\'\i n (\cite{Barrado03}), such strong emission is
consistent with disc accretion phenomena. Other cluster objects show
weak to moderate H$\alpha$ emission lines.

Lamm et al$.$ (2004) have recently detected a strong correlation
between photometric variability in periodic variables and H$\alpha$
emission in a sample of pre-main sequence stars in the young
(2--4\,Myr) open cluster NGC 2264.  However, no evidences for a
correlation for the irregular variables have been found.

\subsection{Near-infrared excess \label{kexcess}}

   \begin{figure}
   \centering
   \includegraphics[width=0.5\textwidth]{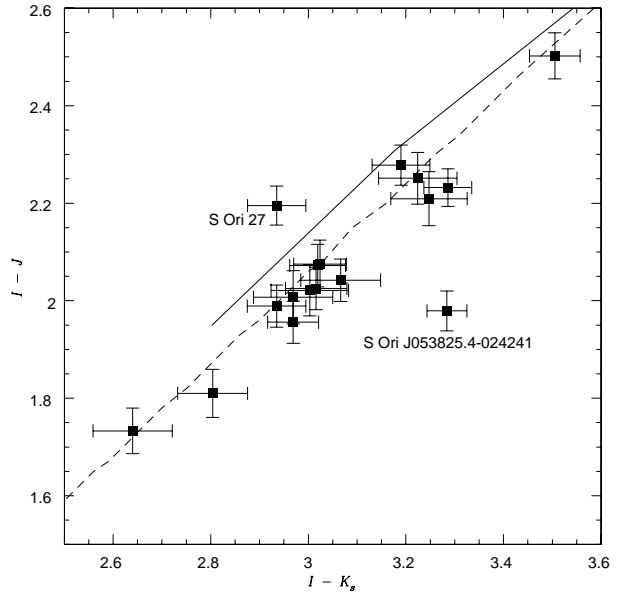}
      \caption{$I-J$ vs.\ $I-K_{\rm s}$ colour--colour diagram.
      All the  brown-dwarf candidates studied with 2MASS $JHK_{\rm s}$ photometry
      data with precision better than 0.1\,mag are shown.
      The isochrones are described in the caption of Fig. \ref{IIJ}.}  
         \label{fso_IJ_IKs}
   \end{figure}
   \begin{figure}
   \centering
   \includegraphics[width=0.5\textwidth]{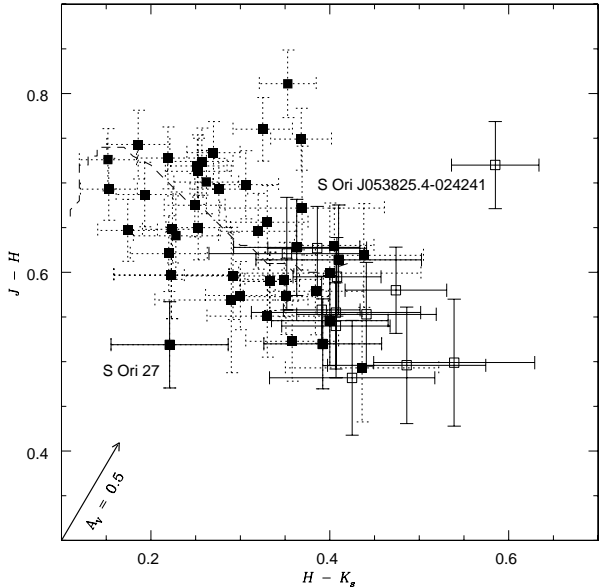}
      \caption{$J-H$ vs.\ $H-K_{\rm s}$ colour--colour diagram.
      Small filled squares with dotted error bars are for
      spectroscopically confirmed objects of the $\sigma$ Orionis cluster.
      Open squares with solid error bars denote the objects studied in
      this paper.
      Spectroscopically confirmed objects are plotted as filled squares.
      The dashed line is the 3 Myr colour--colour isochrone for substellar
      objects, based on  Chabrier et al. (2000) models.
      Late-K and early-M stars are located in the upper left corner of the
      figure, while the mid- and late-M dwarfs are close to the center.}  
         \label{fso_JH_JKs}
   \end{figure}

We have used the 2MASS All-Sky Catalog of Point Sources (Cutri et al$.$\
2003) to study the presence of any near-infrared excesses in our
targets that could be an indication of warm surrounding discs.  In
Fig.~\ref{fso_IJ_IKs} is shown the $I-J$ vs.\ $I-K_{\rm s}$ diagram
for 17 of the brown-dwarf candidates studied, with 2MASS $JHK_{\rm s}$
photometry more precise than 0.1\,mag (i.e.\ targets brighter than $I
=$17.5\,mag, approximately).  The solid and dashed lines are the
3\,Myr colour--colour isochrones for substellar objects, based on
models by the Lyon group (Baraffe et al$.$ 2002 and Chabrier et
al. 2000, respectively).  All the substellar candidates except two lie
in the dashed isochrone within the error bars.  The outliers
are S Ori 27 and S Ori J053825.4$-$024241.  S Ori 27 is an M6.5 $\pm$
0.5 brown dwarf with an estimated mass of $\sim$40\,$M_{\rm Jup}$.
According to its $I-J$ colour, S Ori J053825.4$-$024241 may be an
$\sim$M6-type brown dwarf for which optial spectroscopy is not
available.  While S Ori 27 seems to move towards bluer colours, S Ori
J053825.4$-$024241 shows a clear reddening that suggests an excess of
infrared emission at 2.2\,$\mu$m or, alternatively, it may be affected
by interstellar extinction.  These objects do not lie close to the
cluster sequence in the $J-H$ vs.  $H-K_{\rm s}$ diagram, as shown in
Fig. \ref{fso_JH_JKs} where, again, only 2MASS $JHK_{\rm s}$ data more
precise than 0.1\,mag have been plotted.

Recently, Oliveira et al$.$ (2003) have obtained $L'$ photometry of
$\sigma$ Orionis stars with late K and early M spectral types,
concluding that about 50\%~of the stellar cluster population show a
considerable infrared excess at 3.8\,$\mu$m.  About half of the
$\sigma$ Orionis stars harbour dusty envelopes that may occur in the
form of discs.  We note that the frequency of strong H$\alpha$
emissions in $\sigma$ Orionis stars and brown-dwarf members (Zapatero
Osorio et al$.$ 2002a, 2002b; Barrado y Navascu\'es 2003; Muzerolle et
al. 2003) is not as high as the frequency of $L'$-band infrared
excesses.  Furthermore, the rate of $K$-band infrared excesses is
rather low as compared to the $L'$-band (Oliveira et al$.$ 2002).
Mid-infrared excesses are better disc indicators.

Evidence for accretion discs, dusty envelopes and the existence of a T
Tauri-like phase in young brown dwarfs has been claimed for other
similarly young clusters and star-forming regions, such as $\rho$
Ophiuchi (Wilking, Greene \& Meyer 1999; Testi et al$.$ 2002; Natta et
al. 2002), the Trapezium (Muench et al$.$ 2001), Chamaeleon I (Natta \&
Testi 2001; Persi et al$.$ 2001; Jayawardhana, Mohanty \& Basri 2002), R
Coronae Australis (Fern\'andez \& Comer\'on 2001, Barrado y
Navascu\'es, Mohanty \& Jayawardhana 2003), IC 348 (Jayawardhana,
Mohanty \& Basri 2003), Lupus 3 (Comer\'on et al$.$ 2003), and TW Hydrae
(Mohanty, Jayawardhana \& Barrado y Navascu\'es 2003).

\section{Criteria of variability}

   \begin{figure*}
   \centering
   \includegraphics[width=1.0\textwidth]{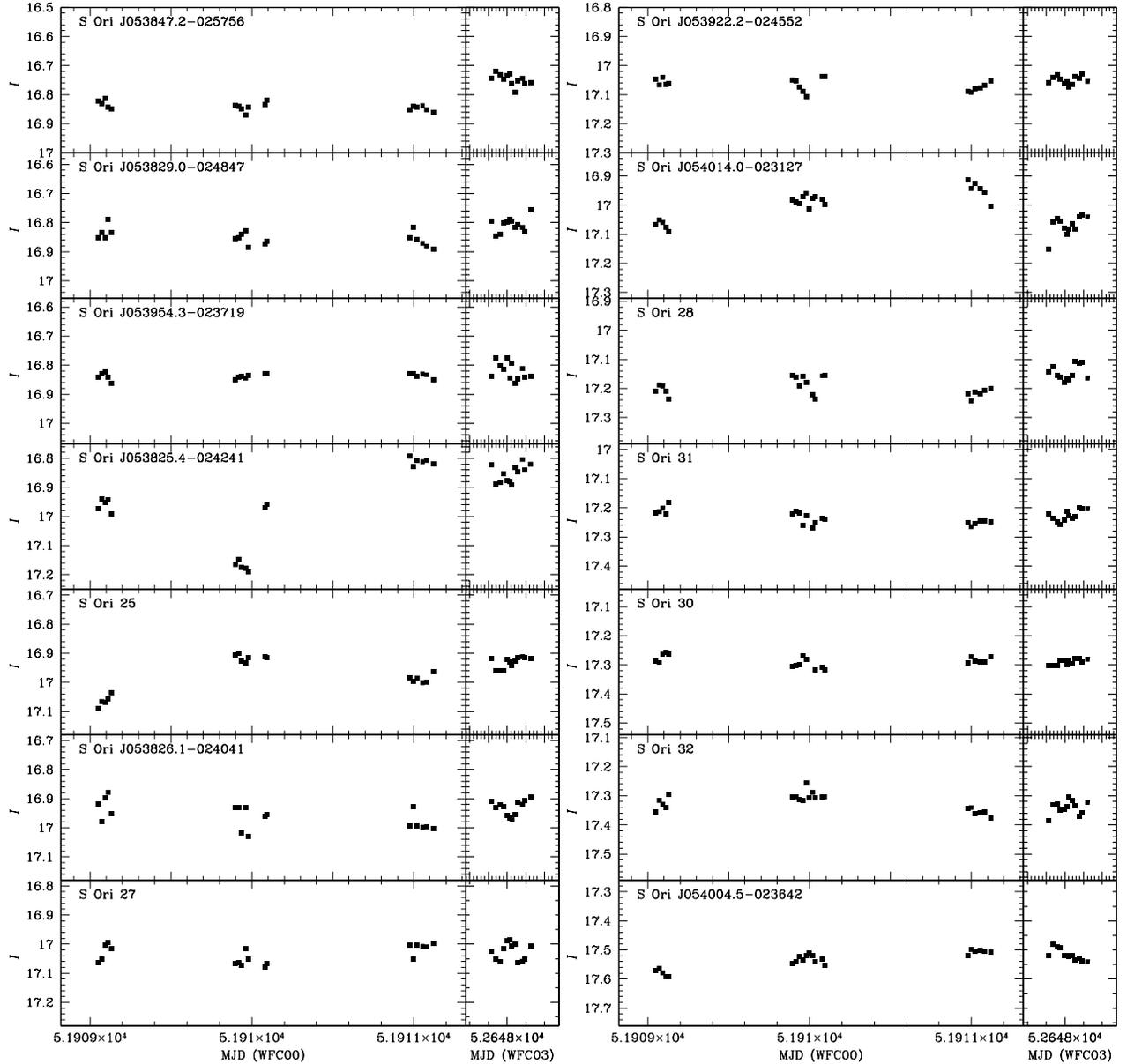}
      \caption{Light curves of the fourteen brightest objects in the final
      sample.}   
         \label{fso_lightone}
   \end{figure*}
   \begin{figure*}
   \centering
   \includegraphics[width=1.0\textwidth]{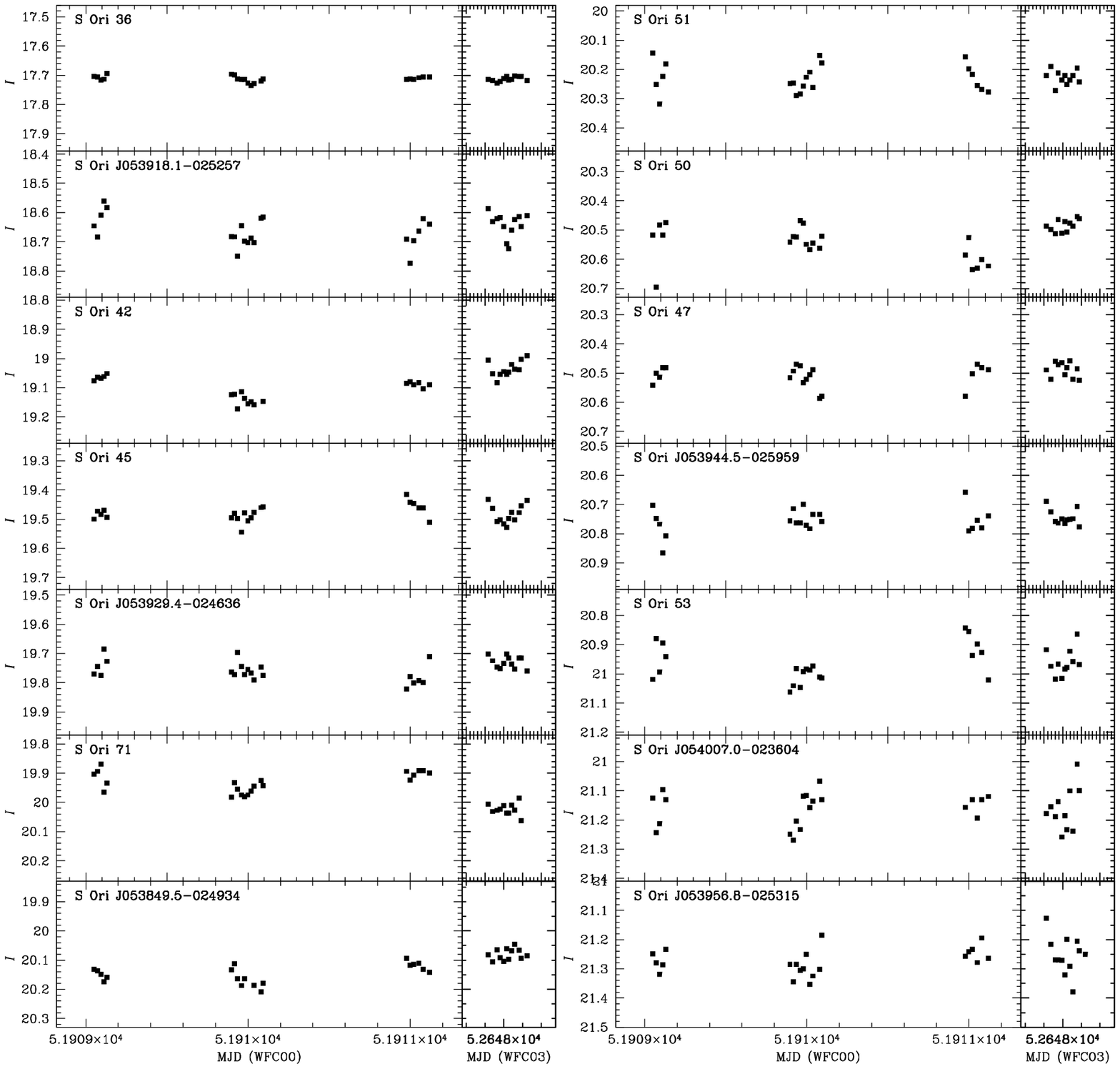}
      \caption{Light curves of the fourteen faintest objects in the final
      sample.}   
         \label{fso_lighttwo}
   \end{figure*}

The scatter of the data points in a light curve represents a measure
of the photometric variability, or at least sets an upper limit to the
variability.  The standard deviation, $\sigma(I)$, of the light curve
is the most used variability indicator, although other parameters have
been also proposed: the peak-to-peak amplitude, the average of the
absolute relative magnitudes, the ``range'' (the difference in median
values of the upper 15\,\% and the lower 15\,\% of measurements),
variability and median statistics, as the $\chi^2$ or the
$\tilde{\eta}$, and other indices or tests as the Kolmogorov--Smirnov
test.  Some of the parameters need posterior Monte Carlo
simulations to quantify the likelihood of variability.  Discussions
about some variability indicators can be found in Stetson (1996),
Bailer-Jones \& Mundt (1999), Herbst et al$.$ (2002) and Enoch et
al. (2003). Here, we will define a few criteria of variability
according to different timescales. 
We will compare the photometry
of our $\sigma$ Orionis sample with the data of field sources of
similar brightness. For consistency purposes, all of the light curves
are treated in the same manner.

We define $\sigma(I)$ as the standard deviation of the light
curves for each night.  The $\sigma(I)$ vs$.$ $I$ diagram
shown in Fig. \ref{IsI00_2} illustrates the strong dependency of the
photometric standard deviation on the target apparent magnitude.
The lower envelope of $\sigma(I)$ is roughly flat at the level of
5\,mmag in the $I \sim$17.0--19.5\,mag range, while it increases
exponentially above $\sim$20.5\,mag because of the Poissonian
photon-count error.  On the other hand, the scatter also increases
below $\sim$16.5\,mag because of saturation and non-linear effects in
bright sources.  We note that to compute the $\sigma(I)$, three $I$
values in the WFC00 epoch and one value in the WFC03 were rejected for
all the objects, as the measured point spread functions (PSF) in the
corresponding images were far from the average PSF for each epoch.

The three brightest brown dwarfs in the sample of 32 $\sigma$ Orionis
objects ($I <$16.7\,mag) will not be considered in the discussion of
Sect.~5 because a significant fraction of the light curve points was
recorded with counts in the non-linear regime of the detector.
Likewise, the light curve of S Ori J053858.6$-$025228, a substellar
object fainter than $I =$ 22\,mag, was discarded because of the large
Poissonian error.  For the remaining 28 brown dwarfs, what we call the
{\it final sample}, the photometry is reliable.

The light curves of our final sample are displayed in Figures
\ref{fso_lightone} and \ref{fso_lighttwo}.  For comparison purposes,
we have used a common display: vertical height has been fixed to
0.5\,mag and the temporal scale is constant along the horizontal
axis. The Modified Julian Date (MJD = JD $- 2400000.5$) is indicated
below, and the two epochs are shown.  We have not taken into account
the heliocentric correction, as the maximum difference between JD and
HJD (Heliocentric Julian Date) is of the order of only a few seconds
over a period of three nights.  The reference star light curves are
similar to that of S Ori 36.

\subsection{Short timescales}
\label{Short_timescales}

   \begin{figure}
   \centering
   \includegraphics[width=0.5\textwidth]{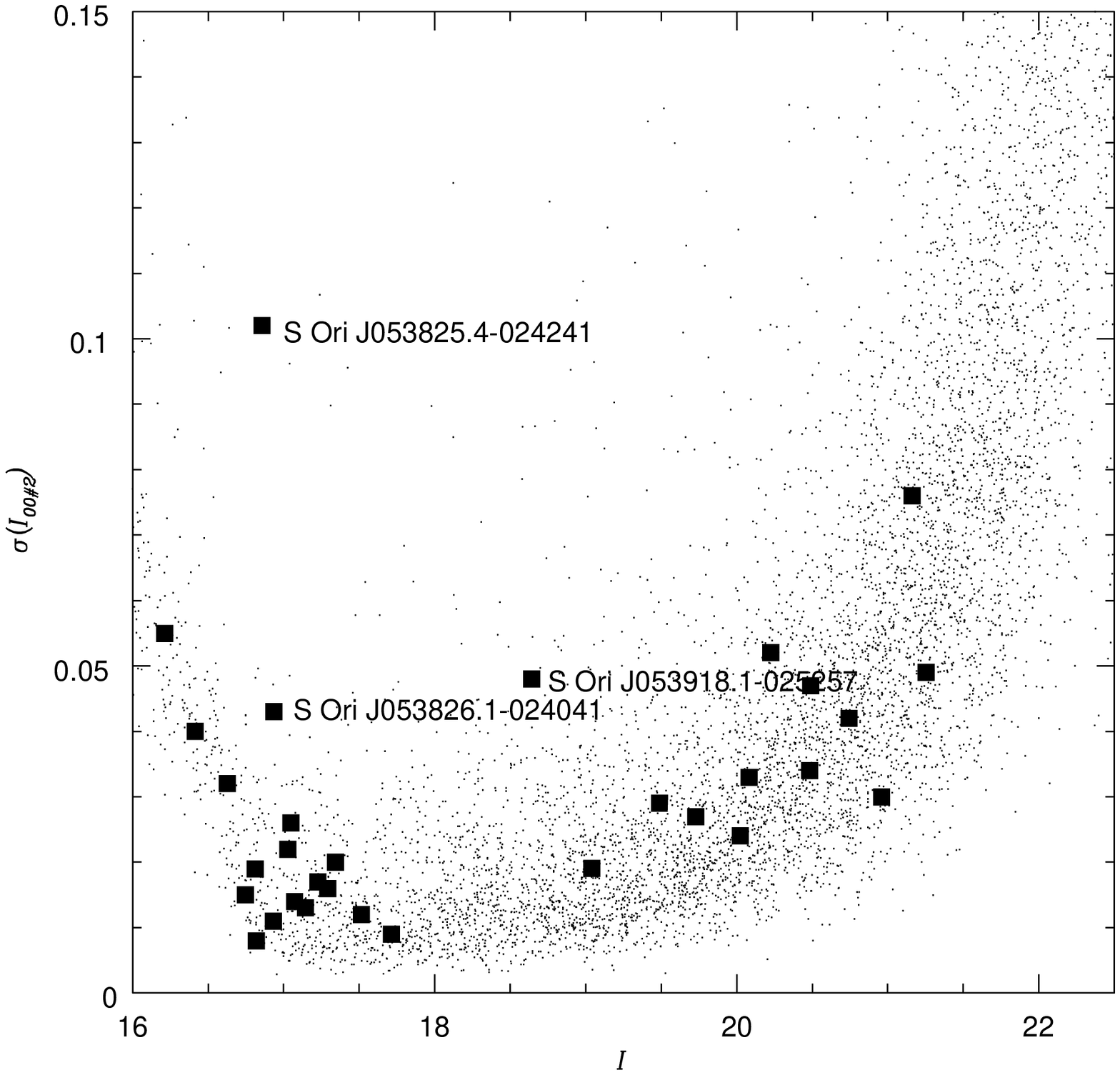}
      \caption{Standard deviation of the differential light curves versus the
      	$I$ magnitudes for each target (small dots) in the second WFC00
	night. 
      	Only objects with $\sigma(I) <$ 0.15\,mag are shown.
	Mean photometric {\sc iraf} errors are below 5, 15 and 100\,mmag
	for objects brighter than $I$=18, 20 and 22, respectively. 
      	Filled squares denote the brown-dwarf candidates.
	Short-term variable brown dwarfs (on the basis of the definition of
	$\varpi_{ST}$) are labelled.} 
         \label{IsI00_2}
   \end{figure}

In order to find photometric variability on very short timescales (in
a series of consecutive exposures, i.e., minutes to hours), we
compared the $\sigma(I)$ of the light curves of the targets with those
of field stars of similar magnitude. For each target we defined a
short-term variability parameter, $\varpi_{\rm ST}$, given by the
following equation:
\begin{equation} 
   \varpi_{\rm ST} = \frac{N^-}{N^+ + N^-} \,,
   \label{varpi}
\end{equation}
where the subscript ``ST'' stands for short-term, and $N^+$ and $N^-$
are the number of sources in the same magnitude bin with $\sigma(I)$
larger and lower, respectively, than the $\sigma(I)$ of the object
under consideration.  High values of $\varpi_{\rm ST}$ imply more
likely variability.  The width of the bin was fixed to 0.2
magnitudes, centred on the $I$ magnitude of the object.  Typically
about one hundred objects with similar brightness and, hence,
with similar photometric error, were considered in the comparison.
Such a narrow bin was selected to minimize the error contribution from
faint sources and sources in the non-linear domain, which show a
larger photometric error that could mask the intrinsic
variability of the target.

We will impose a cut at $\varpi_{\rm ST}$\,=\,0.95 to separate
variable from non-variable sources. Targets with $\varpi_{\rm
ST}$\,$\ge$\,0.95 are considered to be very likely short-term variable
brown dwarfs in our sample. This criterion demands that for a
target brown dwarf to be detected as variable it must be significantly
more variable than field objects in the same magnitude bin. 
Statistically, the light curves of the variable sources
display photometric amplitudes that are among the 5\,\%~largest
amplitudes for their magnitude bins.  This is as if there is only
a 5\%~chance that a brown dwarf which is not truly variable is
incorrectly classified as variable (i.e. 5\%~chance of a false positive
detection). We note that the distribution of $\varpi_{\rm ST}$ is
rather flat between 0 and 1 within a given magnitude bin. However, the
distribution of $\sigma(I)$ is asymmetric: most of the data points
fall close to the mean photometric {\sc iraf} errors, while only a
small percentage lies far from this location (see
Fig. \ref{IsI00_2}).

\subsection{Intermediate timescales}
\label{Intermediate_timescales}

   \begin{figure}
   \centering
   \includegraphics[width=0.5\textwidth]{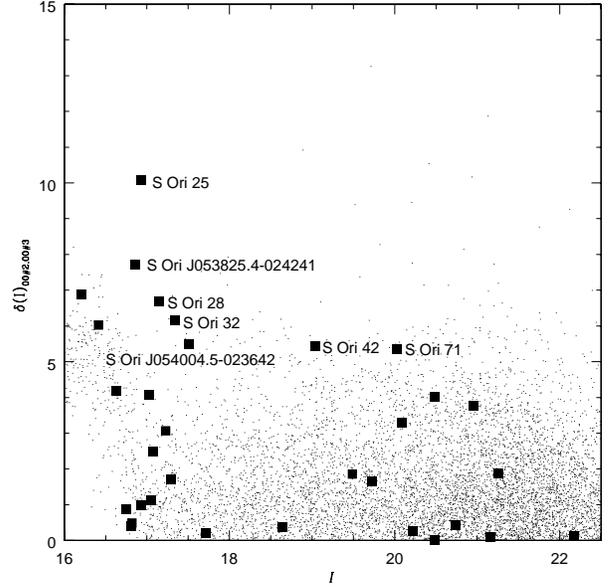}
      \caption{Dimensionless absolute differences of mean magnitudes
      	between the second and third WFC00 nights vs$.$ apparent $I$ magnitudes.
      	Field objects are plotted as small dots, while filled squares denote
	the brown-dwarf candidates.
	Names of several mid-term variable targets are given.}
         \label{newerror_MT}
   \end{figure}

To detect variability in a target from one night to another (hereafter
mid-term variability, i.e., scale of a few days), we compared the
averaged magnitudes (of both targets and field sources),
$\overline{I_{i}}$, computed for each night, $i$. We list the mean
magnitudes of our sample in Table~\ref{nameIsigmaI}. The comparison is
performed by obtaining the absolute value of $\overline{I_{i}} -
\overline{I_{j}}$ per object, where $i$ and $j$ represent two
different nights. In order to avoid the bias introduced by
short-term photometric variability, we have divided the differences
by the standard error of the mean magnitudes.
Hence, we have defined a dimensionless absolute magnitude difference,
$\delta(I)_{ij}$, as:
\begin{equation}
  \delta(I)_{ij} = \frac{|\overline{I_{i}} -
  \overline{I_{j}}|}{\sqrt{\frac{\sigma(I_{i})^2}{n_{i}} +
  \frac{\sigma(I_{j})^2}{n_{j}}}} 
  \label{delta}
\end{equation}
where $n$ stands for the number of photometric data points per
night.  
By using mean $I$-band magnitudes per
night we are increasing the sensitivity to mid-term variability.

We have selected likely mid-term variables in our sample in a
similar manner as we picked short-term variable brown dwarfs. We have
used the definition of equation~\ref{varpi}, which we now name
$\varpi_{\rm MT}$, where the subscript stands for ``mid-term''.
$N^+$ and $N^-$ represent, in this case, the number of sources in the same
magnitude bin with  $\delta(I)_{ij}$ larger and lower, respectively, than the
$\delta(I)_{ij}$ of the object under consideration. 
Figure~\ref{newerror_MT} shows the dimensionless differences 
against apparent magnitudes for the second and third nights of the
WFC00 epoch. Similar diagrams are obtained when considering other
possible combinations, e.g. first and second nights, second and third
nights. We note that mid-term variability can be investigated only in
the WFC00 data. As in the previous analysis of short-term variability,
the bin magnitude is chosen to be 0.2\,mag centred on the $I$
magnitude of each target. According to our criterion, likely mid-term
variable brown dwarfs have $\varpi_{\rm MT} \ge$ 0.95.

\subsection{Long timescales}
\label{Long_timescales}

   \begin{figure}
   \centering
   \includegraphics[width=0.5\textwidth]{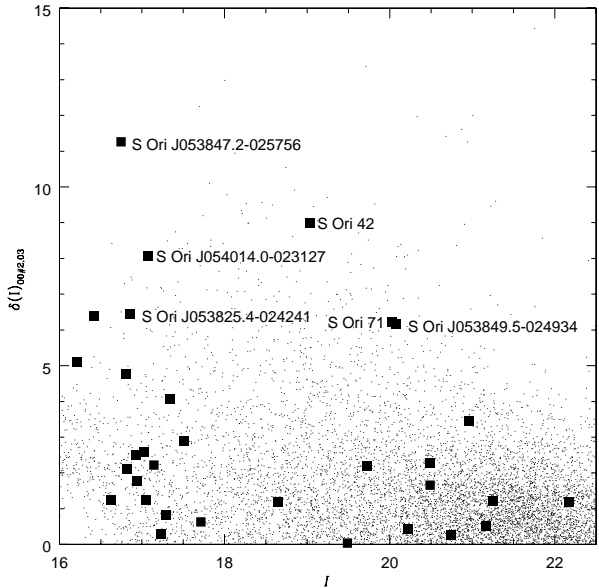}
      \caption{Dimensionless absolute differences of mean magnitudes
        between the second WFC00 night and the WFC03 night vs$.$ apparent
	$I$ magnitudes. 
      	Field objects are plotted as small dots, while filled squares denote
	the brown-dwarf candidates.
	Names of several variable targets are given.}
         \label{newerror_LT}
   \end{figure}

For long-term variability, which takes into account differences
between the first epoch of observations (WFC00) and the second
(WFC03), separated by about 2 years, we applied a criterion
similar to that of mid-term variability. Now, the only WFC03 night is
compared to each of the three WFC00 nights using the dimensionless
magnitude differences, $\delta(I)_{ij}$. 
The parameter that accounts for long-term variability is named $\varpi_{\rm
LT}$.  

Long-term variable $\sigma$ Orionis brown dwarfs in our sample 
have $\varpi_{LT} \ge$ 0.95. The $\varpi_{LT}$ calculated for the
comparison between the first WFC00 night and the WFC03 epoch is
depicted as a function of apparent $I$ magnitude in Fig$.$ \ref{newerror_LT}.
Similar diagrams are obtained for the remaining two WFC00 nights. 
We note that according to the adopted criterion of variability, all mid-term
variables could also be identified as long-term variable sources. This
is so because we have performed comparisons of each WFC00 night to the
WFC03 epoch. Mid-term variable sources might be true
long-term variables but we cannot discriminate easily with our data.   
Hence, we have excluded mid-term variables from our analysis of
long-term variability.

\section{Discussion}

\begin{table*}
     \centering
    	\caption[]{Mean magnitude and standard deviation for the light
	curves of each night$^a$}     
        \label{nameIsigmaI}
        \begin{tabular}{lcccc}
            	\hline
            	\noalign{\smallskip}
Name & 
$\overline{I_{00\#1}} \pm \sigma$  & 
$\overline{I_{00\#2}} \pm \sigma$  & 
$\overline{I_{00\#3}} \pm \sigma$  & 
$\overline{I_{03}} \pm \sigma$  
\\	   
            	\noalign{\smallskip}
            	\hline
            	\noalign{\smallskip}
S Ori J054000.2$-$025159 	& 16.203$\pm$0.012: & 16.34$\pm$0.05:  & 16.17$\pm$0.03:   & 16.21$\pm$0.05:  \\ 
S Ori J053902.1$-$023501 	& 16.51$\pm$0.03:   & 16.58$\pm$0.04:  & 16.47$\pm$0.03:   & 16.40$\pm$0.07:  \\ 
S Ori 16			& 16.578$\pm$0.010: & 16.64$\pm$0.03:  & 16.592$\pm$0.008: & 16.62$\pm$0.03:  \\ 
S Ori J053847.2$-$025756 	& 16.832$\pm$0.015  & 16.842$\pm$0.015 & 16.848$\pm$0.009  & 16.748$\pm$0.020 \\ 
S Ori J053829.0$-$024847	& 16.83$\pm$0.03    & 16.856$\pm$0.019 & 16.86$\pm$0.03    & 16.81$\pm$0.02   \\ 
S Ori J053954.3$-$023719 	& 16.939$\pm$0.015  & 16.838$\pm$0.008 & 16.832$\pm$0.008  & 16.82$\pm$0.03   \\ 
S Ori J053825.4$-$024241 	& 16.96$\pm$0.02    & 17.11$\pm$0.10   & 16.811$\pm$0.012  & 16.85$\pm$0.03   \\ 
S Ori 25 			& 17.063$\pm$0.019  & 16.915$\pm$0.011 & 16.989$\pm$0.015  & 16.932$\pm$0.019 \\ 
S Ori J053826.1$-$024041 	& 16.92$\pm$0.04    & 16.97$\pm$0.04   & 16.99$\pm$0.03    & 16.93$\pm$0.03   \\ 
S Ori 27 			& 17.03$\pm$0.03    & 17.06$\pm$0.02   & 17.01$\pm$0.02    & 17.03$\pm$0.03   \\ 
S Ori J053922.2$-$024552 	& 17.056$\pm$0.012  & 17.06$\pm$0.03   & 17.077$\pm$0.015  & 17.050$\pm$0.014 \\ 
S Ori J054014.0$-$023127 	& 17.069$\pm$0.016  & 16.983$\pm$0.016 & 16.95$\pm$0.03    & 17.07$\pm$0.03   \\ 
S Ori 28			& 17.207$\pm$0.019  & 17.165$\pm$0.013 & 17.217$\pm$0.015  & 17.15$\pm$0.03   \\ 
S Ori 31 	  		& 17.207$\pm$0.016  & 17.230$\pm$0.017 & 17.252$\pm$0.007  & 17.226$\pm$0.019 \\ 
S Ori 30 			& 17.273$\pm$0.016  & 17.297$\pm$0.016 & 17.284$\pm$0.010  & 17.290$\pm$0.010 \\ 
S Ori 32			& 17.33$\pm$0.02    & 17.300$\pm$0.017 & 17.356$\pm$0.013  & 17.34$\pm$0.02   \\ 
S Ori J054004.5$-$023642 	& 17.579$\pm$0.012  & 17.535$\pm$0.013 & 17.506$\pm$0.007  & 17.516$\pm$0.020 \\ 
S Ori 36			& 17.707$\pm$0.009  & 17.710$\pm$0.009 & 17.710$\pm$0.004  & 17.713$\pm$0.008 \\ 
S Ori J053918.1$-$025257 	& 18.62$\pm$0.05    & 18.68$\pm$0.05   & 18.68$\pm$0.05    & 18.64$\pm$0.04   \\ 
S Ori 42 	  		& 19.064$\pm$0.009  & 19.142$\pm$0.019 & 19.088$\pm$0.009  & 19.04$\pm$0.03   \\ 
S Ori 45 	  		& 19.484$\pm$0.013  & 16.49$\pm$0.03   & 19.46$\pm$0.03    & 19.48$\pm$0.03   \\ 
S Ori J053929.4$-$024636 	& 19.74$\pm$0.04    & 19.76$\pm$0.03   & 19.78$\pm$0.04    & 19.73$\pm$0.02   \\ 
S Ori 71 			& 19.91$\pm$0.04    & 19.96$\pm$0.02   & 19.901$\pm$0.012  & 20.02$\pm$0.02   \\ 
S Ori J053849.5$-$024934 	& 20.150$\pm$0.017  & 20.18$\pm$0.04   & 20.118$\pm$0.017  & 20.081$\pm$0.019 \\ 
S Ori 51 			& 20.22$\pm$0.07    & 20.24$\pm$0.04   & 20.23$\pm$0.05    & 20.23$\pm$0.02   \\ 
S Ori 50 			& 20.54$\pm$0.09    & 20.53$\pm$0.03   & 20.60$\pm$0.04    & 20.48$\pm$0.02   \\ 
S Ori 47 			& 20.50$\pm$0.03    & 20.52$\pm$0.04   & 20.52$\pm$0.06    & 20.49$\pm$0.03   \\ 
S Ori J053944.5$-$025959 	& 20.78$\pm$0.06    & 20.75$\pm$0.03   & 20.75$\pm$0.05    & 20.74$\pm$0.03   \\ 
S Ori 53 			& 20.95$\pm$0.06    & 21.01$\pm$0.03   & 20.91$\pm$0.06    & 20.96$\pm$0.04   \\ 
S Ori J054007.0$-$023604	& 21.16$\pm$0.06    & 21.17$\pm$0.07   & 21.18$\pm$0.08    & 21.16$\pm$0.07   \\ 
S Ori J053956.8$-$025315 	& 21.27$\pm$0.04    & 21.29$\pm$0.05   & 21.24$\pm$0.03    & 21.25$\pm$0.07   \\ 
S Ori J053858.6$-$025228	& 21.91$\pm$0.20:   & 22.23$\pm$0.18:  & 22.3$\pm$0.2:     & 22.17$\pm$0.11:  \\ 
        	\noalign{\smallskip}
            	\hline
         \end{tabular}
	\begin{list}{}{}
	\item [$^a$] 00\#1: first observing night of WFC00; 
	00\#2: second observing night of WFC00;
	00\#3: third observing night of WFC00;
	03: observing night of WFC03
	\end{list}
\end{table*}

We provide in Table~\ref{varpi_table} the variability parameters
obtained for the three different timescales and for our target sample.

\begin{table*}
     \centering
    	\caption[]{Long-, mid- and short- term variability parameters of
	the final sample.
	See Section 4 for the definition of each $\varpi$} 
        \label{varpi_table}
        \begin{tabular}{lcccccccccc}
            	\hline
            	\noalign{\smallskip}
Name & 
$\varpi_{\rm LT}$ &
$\varpi_{\rm LT}$ &
$\varpi_{\rm LT}$ &
$\varpi_{\rm MT}$ &
$\varpi_{\rm MT}$ &
$\varpi_{\rm MT}$ &
$\varpi_{\rm ST}$ &
$\varpi_{\rm ST}$ &
$\varpi_{\rm ST}$ &
$\varpi_{\rm ST}$ 
\\	   
            	\noalign{\smallskip}
& 
$_{00\#1-03}$ & 
$_{00\#2-03}$ & 
$_{00\#3-03}$ & 
$_{00\#1-00\#2}$ & 
$_{00\#1-00\#3}$ & 
$_{00\#2-00\#3}$ & 
$_{00\#1}$ & 
$_{00\#2}$ & 
$_{00\#3}$ & 
$_{03}$  
\\	   
            	\noalign{\smallskip}
            	\hline
            	\noalign{\smallskip}
S Ori J053847.2$-$025756       & 0.93 & 0.97 & 0.95 & 0.2  & 0.5  & 0.1  & 0.5  & 0.2  & 0.4  & 0.4  \\ 
S Ori J053829.0$-$024847       & 0.6  & 0.9  & 0.8  & 0.5  & 0.6  & 0.1  & 0.8  & 0.5  & 0.9  & 0.5  \\ 
S Ori J053954.3$-$023719       & 0.6  & 0.5  & 0.4  & 0.0  & 0.1  & 0.0  & 0.5  & 0.1  & 0.3  & 0.8  \\ 
S Ori J053825.4$-$024241       & 0.96 & 1.00 & 0.8  & 0.98 & 0.99 & 0.99 & 0.6  & 0.98 & 0.4  & 0.8  \\ 
S Ori 25		       & 0.97 & 0.6  & 0.9  & 0.97 & 0.94 & 0.94 & 0.6  & 0.4  & 0.6  & 0.5  \\ 
S Ori J053826.1$-$024041       & 0.4  & 0.8  & 0.8  & 0.8  & 0.91 & 0.6  & 0.94 & 0.95 & 0.9  & 0.6  \\ 
S Ori 27		       & 0.1  & 0.8  & 0.4  & 0.7  & 0.4  & 0.90 & 0.7  & 0.7  & 0.7  & 0.7  \\ 
S Ori J053922.2$-$024552       & 0.3  & 0.5  & 0.6  & 0.2  & 0.6  & 0.6  & 0.3  & 0.5  & 0.2  & 0.4  \\ 
S Ori J054014.0$-$023127       & 0.2  & 0.98 & 0.97 & 0.96 & 0.98 & 0.9  & 0.4  & 0.5  & 0.9  & 0.8  \\ 
S Ori 28		       & 0.93 & 0.6  & 0.92 & 0.8  & 0.3  & 0.95 & 0.6  & 0.5  & 0.6  & 0.7  \\ 
S Ori 31		       & 0.7  & 0.1  & 0.6  & 0.6  & 0.8  & 0.8  & 0.6  & 0.7  & 0.2  & 0.5  \\ 
S Ori 30 		       & 0.6  & 0.2  & 0.2  & 0.6  & 0.3  & 0.6  & 0.5  & 0.6  & 0.3  & 0.2  \\ 
S Ori 32		       & 0.5  & 0.8  & 0.4  & 0.7  & 0.7  & 0.97 & 0.6  & 0.7  & 0.4  & 0.6  \\ 
S Ori J054004.5$-$023642       & 0.94 & 0.7  & 0.3  & 0.8  & 0.95 & 0.90 & 0.3  & 0.5  & 0.2  & 0.5  \\ 
S Ori 36		       & 0.2  & 0.1  & 0.1  & 0.1  & 0.2  & 0.0  & 0.2  & 0.3  & 0.0  & 0.1  \\ 
S Ori J053918.1$-$025257       & 0.7  & 0.7  & 0.7  & 0.9  & 0.94 & 0.4  & 0.95 & 0.96 & 0.93 & 0.9  \\ 
S Ori 42		       & 0.6  & 0.98 & 0.8  & 0.9  & 0.5  & 0.8  & 0.1  & 0.97$^*$ & 0.1 & 0.5 \\ 
S Ori 45		       & 0.1  & 0.0  & 0.6  & 0.1  & 0.6  & 0.7  & 0.1  & 0.7  & 0.7  & 0.6  \\ 
S Ori J053929.4$-$024636       & 0.3  & 0.6  & 0.8  & 0.3  & 0.7  & 0.6  & 0.7  & 0.7  & 0.7  & 0.3  \\ 
S Ori 71		       & 0.9  & 0.9  & 0.9  & 0.6  & 0.2  & 0.7  & 0.5  & 0.4  & 0.3  & 0.2  \\ 
S Ori J053849.5$-$024934       & 0.8  & 0.94 & 0.5  & 0.3  & 0.5  & 0.7  & 0.1  & 0.6  & 0.1  & 0.2  \\ 
S Ori 51		       & 0.1  & 0.2  & 0.0  & 0.2  & 0.1  & 0.1  & 0.8  & 0.8  & 0.5  & 0.2  \\ 
S Ori 50		       & 0.7  & 0.6  & 0.8  & 0.3  & 0.7  & 0.8  & 0.90 & 0.4  & 0.3  & 0.1  \\ 
S Ori 47		       & 0.2  & 0.6  & 0.4  & 0.3  & 0.3  & 0.0  & 0.1  & 0.7  & 0.6  & 0.1  \\ 
S Ori J053944.5$-$025959       & 0.4  & 0.1  & 0.1  & 0.4  & 0.4  & 0.1  & 0.6  & 0.4  & 0.4  & 0.1  \\ 
S Ori 53		       & 0.2  & 0.8  & 0.4  & 0.6  & 0.4  & 0.7  & 0.4  & 0.1  & 0.4  & 0.3  \\ 
S Ori J054007.0$-$023604       & 0.0  & 0.3  & 0.1  & 0.2  & 0.2  & 0.0  & 0.3  & 0.7  & 0.5  & 0.7  \\ 
S Ori J053956.8$-$025315       & 0.2  & 0.4  & 0.1  & 0.3  & 0.3  & 0.3  & 0.1  & 0.2  & 0.2  & 0.5  \\ 
        	\noalign{\smallskip}
            	\hline
         \end{tabular}
	\begin{list}{}{}
	\item [$^*$] The S Ori 42 light curve was affected by a crossing
	artificial satellite path.
	\end{list}
\end{table*}

\subsection{Long-term variable brown dwarfs}

Two brown dwarfs, namely {S Ori J053847.2$-$025756} and S
Ori 42, vary significantly ($\varpi_{LT} \ge$ 0.95) from the WFC00 to
the WFC03 epoch.  The amplitude of their light curves is given in
Table~\ref{variable}.  These objects represent about 7\,\%~of
the brown-dwarf final sample.  Because our observations are not
equally sensitive to $I$-band long-term variability as they are to
mid- and short-term photometric variability, this percentage
represents a lower limit to the true frequency of long-term variable
young brown dwarfs.

The exact temporal scale of the observed $I$-band long-term
variability, which may be between several days and years, is unknown
to us.  This kind of variability could be associated with the
long-term evolution of surface features, such as dust clouds or
solar-like spots, or with the presence of an accretion disc or a close
interactive companion.  In the case of S Ori J053847.2$-$025756,
spectral data are lacking, and its H$\alpha$ emission is unknown.
Given its relatively high effective temperature ($\sim$2800\,K,
obtained from its $I-J$ colour), the formation of magnetically-driven
spots or the presence of an accretion disc may be favoured over dust
condensation (Mohanty et al$.$ 2002; Joergens et al$.$ 2003).  
The brown dwarf S\,Ori\,42 is $\sim$2\,mag fainter and cooler, with an
estimated effective temperature at around 2500\,K.  Dust condensation
is expected at these cool temperatures; however, the notably high
H$\alpha$ emission of S\,Ori\,42 (Table~\ref{spectrum}) supports the
scenarios involving mass infalling.

Among our final list of brown dwarfs, S\,Ori\,71 is the strongest
H$\alpha$ emitter (Table~\ref{spectrum}). This source is a
non-variable object according to our criteria. Nevertheless, the three
$\varpi_{LT}$ are quite close to the 0.95 limit, suggesting that it
may be a real long-term variable with an amplitude of
$\sim$0.13\,mag. S\,Ori\,71 is quite faint, and relatively large
photometric errors have likely prevented us from detecting a clear
magnitude variation.  The persistent and strong H$\alpha$ emission of
S Ori 71 suggests that, for this particular object, the emission is
probably linked to the presence of an accretion disc or to a close
interactive companion (Barrado y Navascu\'es et al$.$ 2002).

\subsection{Mid-term variable brown dwarfs}

Six young brown dwarf candidates show significant night-to-night
$I$-band variations during the WFC00 run.  They display at least one
of the three $\varpi_{\rm MT}$ values larger than 0.95.  The amplitude
of their light curves is provided in Table~\ref{variable}. These 
six mid-term variable sources represent about 21\,\% of our
final sample.  Possible causes that lead to mid-term photometric
variability are: variable obscuration by circumsubstellar dust,
heterogeneous distribution of hot or cool spots that cover a large
percentage of the substellar surface, dust condensates in the
atmosphere, variation in accretion rate and transits of faint companions
in close orbits. 
We highlight two cases: S Ori J053825.4$-$024241 and S Ori 25.

S Ori J053825.4$-$024241 is the most variable brown dwarf candidate in
our final sample. The amplitude of the light curve, as measured from
peak-to-peak, is 0.36\,mag. As indicated in Section
\ref{kexcess}, the 2MASS $K_{\rm s}$-band data of S Ori
J053825.4$-$024241 show significant near-infrared excess, which may be
indicative of the presence of a surrounding disc, from which the
central object may be accreting. Hence, the observed $I$-band
variability could be related to episodes of mass accretion or to
eclipses caused by inner portions of the disc, as those presumably
observed in the pre-main-sequence object KH 15D (Hamilton et al$.$
2001). Herbst et al$.$ (2002) stated that a correlation between
near-infrared excesses and large photometric variability in low-mass T
Tauri stars exists.  S Ori J053825.4$-$024241 possibly represents one
of the first examples of a similar correlation among young brown
dwarfs (see also those presented by Carpenter et al$.$ (2001) in the
Chamaeleon I molecular cloud).  However, we note that the connection
between near-infrared excesses and the existence of disc accretion is
not so straight forward.

Optical spectra are also available for S Ori 25.  They
show strong H$\alpha$ emission, with a width of more than
100\,km\,s$^{-1}$ at 10\,\% of peak intensity (see references given in
Table \ref{spectrum}).  The H$\alpha$ emission of S Ori 25 is
remarkably stable on a timescale of years.  In addition, this object
appears to show a modulation in the light curve with a period of
$\sim$40\,h, as we will discuss below.

The other four mid-term variable brown dwarf candidates are S Ori
J054014.0$-$023127, S Ori 28, S Ori 32 and S Ori J054004.5$-$023642.  
Spectroscopy is not available for any of them. However, we note that
S Ori J054014.0$-$023127, for which we determine a photometric $I$-band
amplitude of $\sim$0.12\,mag, appears quite overluminous in the
colour-magnitude diagram of Fig$.$ \ref{IIJ}.

\subsection{Short-term variable brown dwarfs}

Three $\sigma$ Orionis brown dwarf candidates are classifed as
very likely short-term variables according to our $\varpi_{\rm ST}$
criterion.  They are S Ori J053825.4$-$024241, S Ori
J053826.1$-$024041 and S Ori J053918.1$-$025257. Additionally,
although the $\varpi_{\rm ST}$ values of S Ori 27 and S Ori 28 are
below the $\varpi_{\rm ST}$ = 0.95 boundary, we have measured
reliable peaks in the periodograms of their light curves that also
suggest classifying them as periodic short-term variable brown dwarfs (see
Sect.~\ref{periods}).  The amplitudes of the light curves are given
in Table~\ref{variable}.

The strong intra-night photometric variations observed in S Ori
J053826.1$-$024041 and S Ori J053918.1$-$025257 lasts for three or
more days.  Both sources are apparently stochastic photometric
variables with peak-to-peak amplitudes of $\sim$0.10--0.15\,mag on
timescales as short as the time interval between consecutive exposures
($\sim$20\,min).  We cannot classify these two objects as mid-term
variables according to our criteria. However, S Ori
J053825.4$-$024241, which has an infrared excess emission at
2.2\,$\mu$m, complies with our criteria of mid-term
variability. 

In principle, short-term variability is more likely to be related to
fast variations on the surface of an object than in a disc, where the
relevant timescales are likely to be of the order of the revolution
period of its particles. However, the rotational modulation scenario
requires very fast spin periods, as short as one or two hours.  In
contrast, intense vertical motions of convective cells (bubbling)
caused by turbulent convection could modify the photospheric coverage
of dust clouds or dark magnetic spots on very short timescales.  This
is supported by the greater importance of the internal heat source
during the early stages of evolution of substellar objects, which
pushes warm material from the interior to the photosphere.  Mass
infalling from a disc and formation of a hot spot could also explain
the sudden high-amplitude brightening (by more than 200\,mmag) of the unsteady
variable S Ori J053825.4$-$024241 during the second WFC00 night.

\subsection{A search for periodicities \label{periods}}

   \begin{figure}
   \centering
   \includegraphics[width=0.5\textwidth]{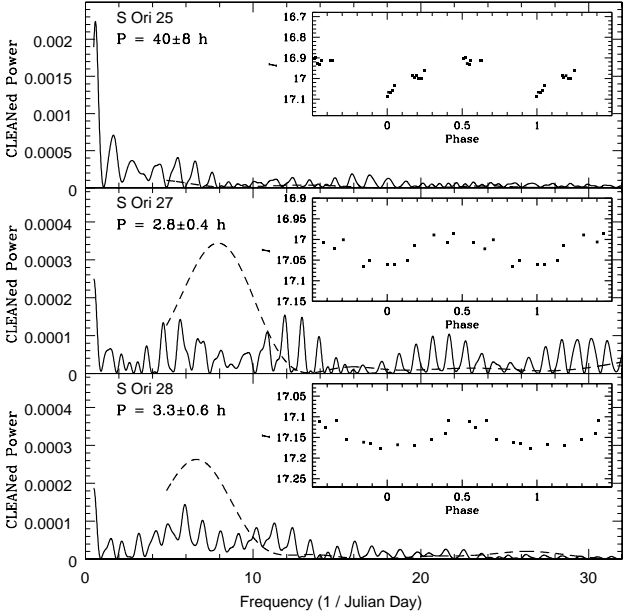}
      \caption{WFC00 (solid line) and WFC03 (dashed line) light curve
      periodograms of S Ori 25, S Ori 27 and S Ori 28 (from top to bottom).
      Insets show the phased light curves associated with each peak of
      the periodogram (WFC00, WFC03 and WFC03, respectively).
      Note that the scale of the {\sc clean}ed power and $I$ magnitude and the
      epoch of S Ori 25 are different from S Ori 27 and 28.
      Further details are given in the text.}  
         \label{fso_phase}
   \end{figure}

Time-series analysis has been performed over our final sample of brown
dwarfs, i.e.\ 56 light curves (28 brown-dwarf candidates $\times$ 2
epochs).  We have used {\sc period}, a time-series analysis package
(developed by Vik Dhillon on La Palma) within the {\sc starlink}
software collection, which includes the Lomb-Scargle (Lomb 1976;
Scargle 1982) and the {\sc clean} (Roberts, Leh\'ar \& Dreher 1987)
algorithms.  Frequency search limits were the Nyquist critical
frequency and the inverse of the maximum temporal coverage.  Five
iterations and a gain of 0.1 were used during the {\sc clean} runs.
The periodograms of three targets (namely, S Ori 25, S Ori 27 and S
Ori 28) show significant peaks, which are at least five times higher
than the typical noise of the periodograms of non-variable field
objects of similar brightness.

S Ori 25 is a mid-term variable brown dwarf with a light curve that
seems to rise and fall in the three-night WFC00 run.  The {\sc
clean}ed periodogram in the top box of the Figure \ref{fso_phase}
shows a very powerful peak at 40 $\pm$ 8\,h (0.61 $\pm$
0.13\,day$^{-1}$) and its harmonics.  The peak also appears in the
Lomb--Scargle periodogram.  The WFC03 data are not useful for checking
this detection, as the coverage was only over $\sim$5\,h.  We will
adopt a radius of 0.45\,$\pm$\,0.05\,$R_\odot$\footnote{1\,$R_\odot$ =
6.9599 $\times$ 10$^8$\,m.} for S Ori 25.  The uncertainty in the
radius determination is estimated from the age interval 1--8\,Myr and
the evolutionary tracks of Chabrier et al$.$ (2002).  The projected
rotational velocity, $v \sin{i}$, of 9.4 $\pm$ 1.0\,km\,s$^{-1}$
(measured by Muzerolle et al$.$ 2003) gives a maximum rotational
period of $P/\sin{i}$ = 58 $\pm$ 9\,h for S Ori 25, which is
compatible with our results.  Furthermore, using the measured
rotational period, we can constrain both the inclination angle ($i
=46^{+16}_{-13}$ degrees) and the rotational velocity (14 $\pm$
4\,km\,s$^{-1}$).  The 40\,h period is greater by far than those in
the literature for field brown dwarfs (except for the 238\,h period
detected in 2MASSW J1300425$+$191235 by Gelino et al$.$\ 2002), but
similar to those recently discovered by Joergens et al$.$\ (2003) in
the young Chamaeleon I star-forming region.  We note that while
the duration of the periodic modulation of S Ori 25 appears to be
generally longer than that of field brown dwarfs, its $v \sin{i}$
measurement is similar (Mohanty \& Basri 2003; Bailer-Jones 2004). In
addition, the strong H$\alpha$ emission in S Ori 25 suggests mass
accretion.  A surrounding accretion disc rotationally locked to the
young central object could explain the low angular velocity of S Ori
25.  The angular velocity would be kept roughly constant at a low rate
until the disc dissipates and the locking brakes. A similar picture is
thought to occur in classical T Tauri stars.

The periodograms of S Ori 27 and S Ori 28 display single peaks
at 2.8 $\pm$ 0.4 and 3.3 $\pm$ 0.6\,h, respectively. Folded WFC03
light curves clearly show sinusoidal modulations with amplitudes
of about 30\,mmag, as illustrated in Figure
\ref{fso_phase}. Unfortunately, the WFC03 data cover roughly one
cycle; hence, the determined periods are quite uncertain. The WFC00
light curves do not show periodical variations in time scales of
$\sim$ 3\,h. We recall that S Ori 28 is also a mid-term variable
brown dwarf candidate.

This kind of variability, possibly related to surface rotation, could
be caused by heterogeneities spread over the photosphere (dust or
spots) that evolve over timescales of several hours.  The rapid
evolution of atmospheric features that lead to the detection of
``false'' periods if only pieces of light curves are taken into
account has been previously noticed by Bailer-Jones \& Mundt (2001)
and Gelino et al$.$  (2002).  Analogously to the Type II$p$ classical T
Tauri stars described by Herbst et al$.$\ (1994), hot spots linked
to unsteady accretion or rotation could originate the modulations.
Hot spots do not last for hundreds or thousands of rotations, as
cool spots do.

Other studies of photometric variability of $\sigma$ Orionis cluster
substellar members have been published by Bailer-Jones \& Mundt (2001)
and ZO03.  S Ori 31 and S Ori 45 are the only objects in common with
this work.  Bailer-Jones \& Mundt (2001) claimed a rotation period of
7.5 $\pm$ 0.6\,h for S Ori 31.  Although this timescale is suitably
studied with the WFC00 data, the suggested amplitude of the
photometric variation is quite small, $\sim$10\,mmag.  These authors
also found a peak of 30 $\pm$ 8\,min in the periodogram of S Ori 45, a
$\sim$25 $M_{\rm Jup}$ M8.5 brown dwarf with variable, strong
H$\alpha$ emission.  A comparably short period ($\sim$46\,min) has
also been claimed by ZO03.  However, as noted by those authors, given
the expected radius of this young brown dwarf, such a short period
cannot be associated with rotation because it would imply an extremely
fast rotational velocity inconsistent with hydrostatic equilibrium.

ZO03 also detected a second, more powerful, modulation in their
optical and infrared light curves of S Ori 45 with a period of
2.5--3.6\,h that could be linked to rotation.  The present WFC00
dataset has already been considered by ZO03.  Our new analysis led to
a very similar optical light curve.  We retrieved the claimed period
once the data points around phase 0.5 were excluded, as suggested in
ZO03 after using the information from the infrared
photometry. According to our rather conservative criteria for the
determination of photometric variability, S Ori 45 lies close to the
borderline between variable and non-variable sources. We note that
this is also the case of the periodic brown dwarfs S Ori 27 and S Ori
28. From our work and the data collected from the literature, we
find that the frequency rate of periodic brown dwarfs in our final
sample is around 18\,\%.

\subsection{Final remarks}

   \begin{figure}
   \centering
   \includegraphics[width=0.5\textwidth]{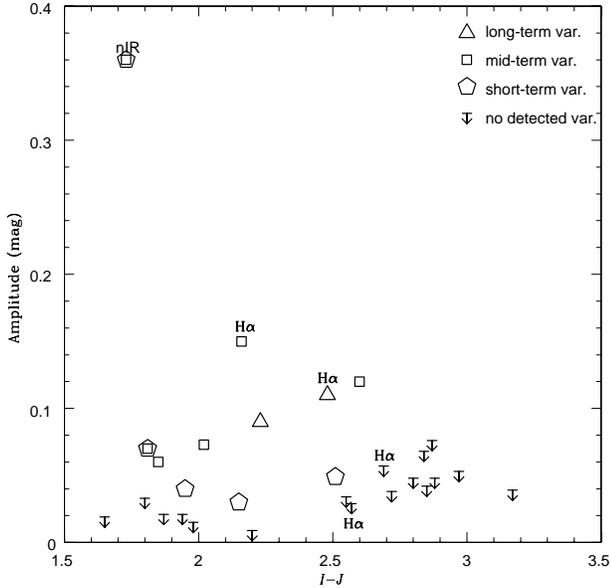}
      \caption{Amplitudes of variability versus $I-J$ colour.
      Open triangles, squares and pentagons denote long-, mid- and short-term
      variables, respectively.
      Upper limits are shown for the rest of the objects in our sample.
      Strong H$\alpha$ emission and near-infrared excess are also indicated.}  
         \label{fso_IJ_mag.ps}
   \end{figure}

Of our final list of twenty eight young brown dwarfs in the
$\sigma$\,Orionis cluster, nine show some kind of optical
photometric variability (short-, mid-, long-term and periodic types) with a
5\,\% chance of a false positive detection,
which represents 39\,\%~of the sample.  
This frequency rate increases up to 46\,\%~if S Ori 27, S Ori 28
(periodic short-term variables with slightly larger false positive detection
probabilities), S Ori 31 and S Ori 45 (periodic brown dwarfs reported in the
literature) are also considered.   
If we only take into account the $\sigma$\,Orionis brown dwarf candidates
brighter than $I$=19.5, for which the photometric errors are
reasonably low, the frequency of variability turns out to be 72\,\%. 
We summarize our results in Table~\ref{variable}, where the abbreviations
LT, MT and ST stand for long-, mid- and short-term variability,
respectively.  If a periodic signal is reported either in the present
study or in the literature, the letter ``P'' is used.  S Ori 27, 28,
31 and 45 are considered as ST variables because of their short
photometric periodicities.  $I$-band amplitudes or scatters of the
light curves and other interesting properties, like strong H$\alpha$
emission and near-infrared excess, are also provided in
Table~\ref{variable}.

In Figure \ref{fso_IJ_mag.ps} we plot maximum photometric amplitudes
for the final sample of objects against the $I-J$ colour, which is an
indicator of temperature and mass.  We use different symbols for the
various types of variability, and downward-point arrows for sources
with no detected variability.  As inferred from the figure, there is
no clear trend between amplitudes and the colour of the brown dwarfs.
S Ori J053825.4$-$024241 shows the largest amplitude, which, as
discussed above, may be associated with the presence of a surrounding
disc.  It is also noteworthy from the figure that the largest $I$-band
variabilities are found in the group of mid- and long-term variable brown
dwarfs.  We find $I$-band photometric variability in $\sigma$
Orionis cluster members with masses in the range 70--25\,M$_{\rm
Jup}$.

\begin{table*}
     \centering
    	\caption[]{Summary of photometric variability}    
        \label{variable}
        \begin{tabular}{lccc}
            	\hline
            	\noalign{\smallskip}
Name & 
Variability & 
Amplitude & 
Properties$^{\mathrm{c}}$  \\	   
            	\noalign{\smallskip}
& 
type$^{\mathrm{a}}$ & 
[mag]$^{\mathrm{b}}$ &
\\	   
            	\noalign{\smallskip}
            	\hline
            	\noalign{\smallskip}
S Ori J053847.2$-$025756       & LT                    & 0.09$\pm$0.02   & -- \\ 
S Ori J053829.0$-$024847       & --                    & $\le$0.03       & -- \\ 
S Ori J053954.3$-$023719       & --                    & $\le$0.02       & -- \\ 
S Ori J053825.4$-$024241       & MT + ST               & 0.36$\pm$0.04   & nIR \\ 
S Ori 25		       & MT (P)                & 0.15$\pm$0.02   & H$\alpha$ \\ 
S Ori J053826.1$-$024041       & ST                    & 0.04            & -- \\ 
S Ori 27		       & ST (P)                & 0.03            & -- \\ 
S Ori J053922.2$-$024552       & --                    & $\le$0.019      & -- \\ 
S Ori J054014.0$-$023127       & MT                    & 0.12$\pm$0.05   & -- \\ 
S Ori 28		       & MT + ST (P)           & 0.07$\pm$0.03   & -- \\ 
S Ori 31		       & ST (P$^{\mathrm{d}}$) & $\le$0.02       & -- \\ 
S Ori 30 		       & --                    & $\le$0.015      & -- \\ 
S Ori 32		       & MT                    & 0.06$\pm$0.02   & -- \\ 
S Ori J054004.5$-$023642       & MT                    & 0.073$\pm$0.014 & -- \\ 
S Ori 36		       & --                    & $\le$0.009      & -- \\ 
S Ori J053918.1$-$025257       & ST                    & 0.05            & -- \\ 
S Ori 42		       & LT                    & 0.11$\pm$0.03   & H$\alpha$ \\ 
S Ori 45		       & ST (P$^{\mathrm{d}}$) & $\le$0.03       & H$\alpha$ \\ 
S Ori J053929.4$-$024636       & --                    & $\le$0.03       & -- \\ 
S Ori 71		       & --                    & $\le$0.06       & H$\alpha$ \\ 
S Ori J053849.5$-$024934       & --                    & $\le$0.05	 & -- \\ 
S Ori 51		       & --                    & $\le$0.04       & -- \\ 
S Ori 50		       & --                    & $\le$0.05   	 & -- \\ 
S Ori 47		       & --                    & $\le$0.04       & -- \\ 
S Ori J053944.5$-$025959       & --                    & $\le$0.04       & -- \\ 
S Ori 53		       & --                    & $\le$0.08       & -- \\ 
S Ori J054007.0$-$023604       & --                    & $\le$0.07       & -- \\ 
S Ori J053956.8$-$025315       & --                    & $\le$0.05       & -- \\ 
        	\noalign{\smallskip}
            	\hline
         \end{tabular}
	\begin{list}{}{}
	\item[$^{\mathrm{a}}$] Variability type. LT: long-term; MT: mid-term;
	ST: short-term; P: periodic; --: no detected variability.
	\item[$^{\mathrm{b}}$] Amplitudes of photometric variability:
	peak-to-peak amplitudes for LT and MT variables and
	$\sigma(I)$ amplitudes for ST and non-variables.
	For the latter, the values must be understood as upper limits on
	variability.
	\item[$^{\mathrm{c}}$] Properties or prominent spectral features. 
	H$\alpha$: strong H$\alpha$ emission; nIR: near-infrared excess.
	\item[$^{\mathrm{d}}$] Periodic according to Bailer-Jones \& Mundt
	(2001) and/or ZO03. 
	\end{list}
\end{table*}

\section{Conclusions and summary}

We have photometrically monitored in the $I$-band a sample of thirty
two young brown-dwarf candidates in the young star cluster $\sigma$
Orionis. Many of them are spectroscopically confirmed as bona fide
$\sigma$ Orionis substellar members. They are $\sim$3\,Myr old, and their
masses span from the substellar limit down to the planetary-mass
borderline. Twenty-eight of these substellar objects have reliable
variability indicators, and their light curves have been analysed in
detail. This is one of the largest samples of young brown
dwarfs ever monitored for photometric variability.

With a 5\,\% chance of a false positive detection, we have found
that nine of our targets show $I$-band variability on various
timescales.  If we include the periodic young brown dwarfs S Ori 27
and S Ori 28 found in our work, and S Ori 31 and S Ori 45 reported in
the literature, the incidence of variable $\sigma$ Orionis brown
dwarfs is around 46\,\%.  Such a frequency is higher than that
(about 33\,\%) found among very low mass field stars and brown dwarfs.
The measured amplitudes of variability are also more than three times
larger in this study.  This different behaviour can be ascribed to the
youth of the $\sigma$ Orionis brown dwarfs.  We classify photometric
variability on the basis of long-, mid- and short-term variations.

\begin{itemize}
\item {\it Long-term photometric variability (year-to-year
variations)}.  Two substellar candidates (a minimum of 7\,\% of the
final sample) show variations in the mean differential magnitudes
measured at two epochs 2.09 years apart, with amplitudes of
0.09--0.11\,mag.  They show neither inter- nor intra-night variations.
\item {\it Mid-term variability (day-to-day variations)}.  Six
targets (21\,\%) appear to be variable from night to night, with
amplitudes of variation in the range 0.06--0.36\,mag.  Three of
them have magnitude variations greater than 0.1\,mag, which may make
these the brown dwarfs with the largest amplitudes detected to date.
S Ori 25 shows a $\sim$40\,h periodic modulation in the light curve,
consistent with the recent determination of the $v \sin{i}$ parameter,
which yields a rotational velocity of 14 $\pm$ 4\,km\,s$^{-1}$.
Magnetic braking due to an accretion disc could account for the long
period.
\item {\it Short-term variability (scale of hours)}.  
Three of our objects apparently display random non-periodic photometric
$I$-band variability on timescales of less than a few hours. 
The amplitudes of the light curves of two of these objects appear to be roughly
steady during at least three consecutive nights with standard deviations at the
0.04--0.05\,mag level. 
The third short-term variable brown dwarf, S Ori J053825.4$-$024241, is
also a mid-term variable and shows the largest variability in our final sample.

With slightly larger false positive detection probabilities, S Ori 27 and S Ori
28 display peaks in the WFC03 periodograms at around 3\,h, which could be
related to fast rotation periods, such as those found 
S
Ori 27 and S Ori 28 additionally show peaks in the WFC03 periodograms
at around $\sim$3\,h, which could be related to fast rotation periods,
such as those found in other young brown dwarfs and evolved ultracool
field dwarfs. These rapid rotations could be explained by the
dissipation of a surrounding disc, or by the evolution of surface
features on timescales of a few rotations.
\end{itemize}

We note the correlation found between large amplitude of variability
and the detection of a near-infrared excess or strong H$\alpha$
emission, which could be an evidence of a substellar accretion disc.
Four out of thirteen brown dwarfs with available spectra show strong
H$\alpha$ emission (i.e.\ pEW(H$\alpha$) $>$ 30\,\AA).  Two of
them (S Ori 25 and S Ori 42) are variable with amplitudes of
$\sim$0.12\,mag and with timescales longer than one day.  
Regarding the other two, S Ori 45 has been independently confirmed as
as photometrically variable by ZO03, and S Ori 71 is suspected to be a
faint long-term variable.  An infrared excess in the $K_{\rm s}$ band
has been detected in the largest-amplitude variable brown dwarf of the
sample, S Ori J053825.4$-$024241.  This is one of the first substellar
objects sharing a correlation between near-infrared excess and large
photometric variability, in analogy to low-mass T Tauri-like stars.

A single variability scenario can hardly explain simultaneously all
the temporal scales involved.  The presence of surrounding accretion
discs, heterogeneous coverage over the photosphere by cool magnetic
spots or fast-evolving dust clouds, mass transfer from or eclipses due
to a companion have been suggested as possible causes of the
variability.  Both long-term, multi-band coverage monitoring and
complete spectroscopic studies are needed to conclude unambiguously
which is the preferred variability scenario in young objects below the
substellar limit.

\begin{acknowledgements}
  	We thank C. A. L. Bailer-Jones for a careful reading of
 	the manuscript and for his suggestions, which have clearly
 	improved this manuscript.  We would like to thank to
 	E. Mart\'{\i}n and D. Barrado y Navascu\'es for helpful
 	comments, and Terry Mahoney for revising the English of the
 	manuscript.  Based on observations made with the Isaac Newton
 	Telescope (INT) operated on the island of La Palma by the
 	Isaac Newton Group in the Spanish Observatorio del Roque de
 	Los Muchachos of the Instituto de Astrof\'{\i}sica de
 	Canarias.  Partial financial support was provided by the
 	Spanish Ministerio de Ciencia y Tecnolog\'{\i}a proyect
 	AYA2001-1657 of the Plan Nacional de Astronom\'{\i}a y
 	Astrof\'{\i}sica.  Part of this work has been supported by the
 	Spanish {\itshape Programa Ram\'on y Cajal}.  This research
 	has made use of the SIMBAD database, operated at CDS,
 	Strasbourg, France.  This publication makes use of data
 	products from the Two Micron All Sky Survey, which is a joint
 	project of the University of Massachusetts and the Infrared
 	Processing and Analysis Center/California Institute of
 	Technology, funded by the National Aeronautics and Space
 	Administration and the National Science Foundation.
\end{acknowledgements}

\end{document}